\documentclass{IEEEtran}
\usepackage[utf8]{inputenc}
\usepackage{times}
\usepackage{color,soul}
\usepackage{graphicx}
\usepackage{epsfig}
\usepackage{amsmath}
\usepackage{amssymb}
\usepackage{amsthm}
\usepackage{amsmath}
\usepackage{array}
\usepackage{multirow}
\usepackage{color}
\usepackage[table]{xcolor}
\usepackage{url}
\usepackage{xcolor}
\usepackage{lipsum}
\usepackage{svg}
\usepackage{wrapfig}
\usepackage{breqn}
\usepackage{multicol}
\usepackage{lastpage}
\usepackage[font=scriptsize]{caption}

\usepackage{fixltx2e}
\usepackage{xcolor}
\usepackage{float}
\usepackage{lipsum}
\usepackage{afterpage}
\usepackage{blindtext}
\makeatletter
\usepackage{adjustbox}
\usepackage{textgreek}
\usepackage{etoolbox}
\usepackage[table]{xcolor}
\usepackage{subcaption}

  \def\vs{{\bf s}} \def\vt{{\bf t}}
   \def\vx{{\bf x}}

\begin{document}

\title{EMAHA-DB1: A New Upper Limb sEMG Dataset for Classification of Activities of Daily Living}
\author{Naveen Kumar Karnam, Anish Chand Turlapaty,~\IEEEmembership{Member,~IEEE}, Shiv Ram Dubey,~\IEEEmembership{Senior Member,~IEEE} and Balakrishna Gokaraju,~\IEEEmembership{Member,~IEEE} 
\thanks{N.K. Karnam and A.C. Turlapaty are with the Biosignal Analysis Lab at  the Indian Institute of Information Technology, Sri City, A.P., India (email: anish.turlapaty@iiits.in).}
\thanks{S.R. Dubey is with the Computer Vision and Biometrics Laboratory at Indian Institute of Information Technology, Allahabad, Prayagraj-211015, U.P., India (email: srdubey@iiita.ac.in).}
\thanks{B. Gokaraju is with the Visualizations and Computing Advanced Research Center (ViCAR) and Department of Computational Data Science an Engineering, North Carolina A and T State University, Greensboro, North Carolina (email: bgokaraju@ncat.edu).}}


\maketitle

\begin{abstract}
In this paper, we present electromyography analysis of human activity - database 1 (EMAHA-DB1), a novel dataset of multi-channel surface electromyography (sEMG) signals to evaluate the activities of daily living (ADL). The dataset is acquired from $25$ able-bodied subjects while performing $22$ activities categorised according to functional arm activity behavioral system (FAABOS) (3 - full hand gestures, 6 - open/close office draw, 8 - grasping and holding of small office objects, 2 - {flexion and extension of finger movements}, 2 - writing and 1 - rest). The sEMG data is measured by a set of five Noraxon Ultium wireless sEMG sensors with Ag/Agcl electrodes placed on a human hand. The dataset is analyzed for hand activity recognition classification performance. The classification is performed using four state-of-the-art machine learning classifiers, including Random Forest (RF), Fine K-Nearest Neighbour (KNN), Ensemble KNN (sKNN) and Support Vector Machine (SVM) with seven combinations of time domain and frequency domain feature sets. The state-of-the-art classification accuracy on five FAABOS categories is $83.21\%$ by using the SVM classifier with the third order polynomial kernel using energy feature and auto regressive feature set ensemble. The classification accuracy on $22$ class hand activities is $75.39\%$ by the same SVM classifier with the log moments in frequency domain (LMF) feature, modified LMF, time domain statistical (TDS) feature, spectral band powers (SBP), channel cross correlation and local binary patterns (LBP) set ensemble. The analysis depicts the technical challenges addressed by the dataset. The developed dataset can be used as a benchmark for various classification methods as well as for sEMG signal analysis corresponding to ADL and for the development of prosthetics and other wearable robotics. 
\end{abstract}


\section{Introduction}
\label{sec:Introduction}

\IEEEPARstart{P}{erforming} hand movements during activities of daily living ({ADL}) \cite{nguyen2016recognition} without any difficulty provides functional independence and {a decent quality of life} \cite{monjazebi2016functional}. However, it is quite difficult to perform simple hand movements for individuals {affected by} the following disorders: upper limb disabilities \cite{world2021brief,narang1986clinical}, disorders related to aging  \cite{overdorp2016combined}, neuromuscular disorders \cite{deenen2015epidemiology}, and stroke related disabilities \cite{chieffo2016noninvasive,kenmuir2015predictors,pandian2013stroke, 9770799}.
Human computer interfaces and human robot interfaces can support the rehabilitation process to recover from the above mentioned disorders. For instance, hand gesture-based interfaces based on computer vision techniques for identifying and classifying gestures are currently under development \cite{mitra2007gesture}. Moreover, many researchers have explored {robotic control using visual} gestures \cite{waldherr2000gesture, yang2007gesture,burger2012two}. However, vision based {control} methods are inadequate to determine the {appropriate control} for actuation and the amount of force exerted by a muscle during action.
One approach to quantify the upper limb activity is to use wearable sensors such as inertial motion sensors (IMUs) including accelerometers, gyroscopes and magnetometers. These sensors are utilised {to  measure and monitor} limb activities, {quantify} muscle motor deficits \cite{9373017}, and {classify} the types of physical activity \cite{karantonis2006implementation,yang2008using}. Although wearable sensors can recognize human activity, they are {deficient} in precise identification of hand gestures, finer finger movements and the amount of muscle strength used to execute the movement \cite{qi2020smartphone}.
 
Alternatively, hand movement classification and the limb control  \cite{bell2008control,farina2014extraction} through surface electromyography (sEMG) signals facilitates the design of prosthetic devices, exoskeleton arms, advanced realistic bio-mechanical models, and rehabilitation therapies \cite{wen2014hand}. In these applications, utilization of multi-modal signals is also very common. In the literature, fusion of the IMU's and sEMG signals  \cite{wolf2013gesture,lu2014hand,jiang2017feasibility} for hand activity classification and estimation of the continuous orientation of the forearm is analyzed. The electroencephalography signals (EEG)) and sEMG signals are also fused to decode the intention of the person. This fusion process can generate better control signals {compared to  a standalone sEMG signal based control}  \cite{8704181, jeong2020multimodal,tryon2021evaluating}. In order to obtain better classification accuracies the features from sEMG signals {can be fused with those from} the vision based image classification network \cite{zandigohar2021multimodal}.
In practice, the multi-modal methods increase the complexity of the hardware as well as software systems, hence they pose difficulty for different real-life applications. Hand movement analysis and classification through standalone sEMG signals is gaining attention  \cite{8007300, jarque2019calibrated,song2022activities, 9855497, 9745921} and is the focus of our current work. 

In this paper, we present electromyography analysis of human activity - database 1 (EMAHA-DB1), a novel sEMG dataset on ADL for the Indian population. Following are the salient features of EMAHA-DB1:
\label{SalientFeat}
\begin{itemize}
    \item There are several sEMG datasets available that include activities such as hand gestures, hand movements, wrist movements, and grasping objects. These datasets are mainly collected for western populations and there is no dataset for ADL from the Indian population. EMAHA-DB1 fills this gap. 
    \item There is a tradition of anthropometric data collection in India \cite{10.2307/40383967}, \cite{majumder2018anthropometry}. For any population, there is an influence of anthropometrics on their kinematics and kinetics \cite{liang2020asian}, \cite{7363534}. EMAHA-DB1 will compliment existing anthropometrics, kinematics and kinetics datasets \cite{jarque2019calibrated}, \cite{7363534}  which will be helpful in conducting upper limb rehabilitation therapies, physiological studies and clinical studies for Indian population.
    
    \item The ADL performance is {analyzed by grouping the actions according to the functional arm activity behavioral observation system} (FAABOS \cite{uswatte2009behavioral}). The functional taxonomy provided by Uswatte \textit{et al.} quantifies group of hand actions based on the behavioral significance.
    
    \item There are publicly available ADL datasets such as the NinaPro \cite{Atzori2014}, the BioPatRec \cite{Ortiz-Catalan2013}, the Ramikushaba \cite{khushaba2014towards} and the UCI Gesture \cite{lobov2018latent} that have not covered a few important ADL categories. The hand activities are usually performed in an experimental set up with a fixed duration for each of the activities, however we have considered different durations for distinct activities to approximate corresponding durations of real time hand movements.
    
    \item The dataset can be used to benchmark classification algorithms or perform statistical studies. The developed dataset consists of a larger number of subjects and a higher number of activity repetitions compared to any other publicly available ADL datasets. 
\end{itemize} 

\begin{figure*}[!t]
\centering
\includegraphics*[width=0.98\textwidth,height=9cm]
{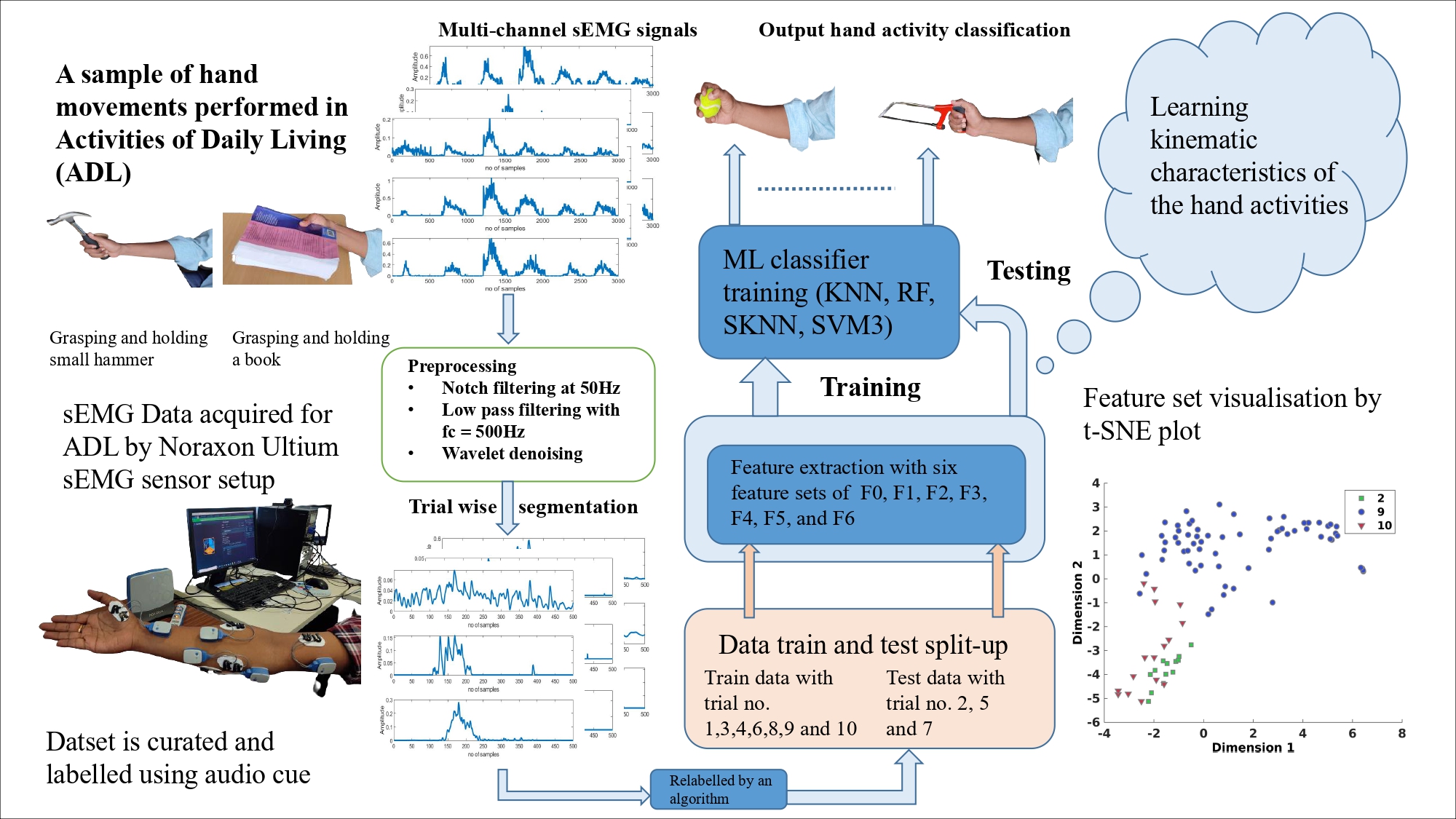} 
\caption{Learning steps from sEMG dataset collection to classification of hand activities}
\label{fig:Block diagram}
\end{figure*}




The main contributions of the paper are:
\begin{enumerate}
    \item In this work, we have carried out muscle activity measurements corresponding to activities of daily living and collected a novel multichannel sEMG data from Indian population. 
    \item The EMAHA-DB1 dataset is organized according to custom FAABOS functional categories to perform analysis using state-of-the-art machine learning classifiers. Specifically the sEMG signals are analyzed to classify into the functional groups as well as individual activities.
    \item We also perform extensive feature analysis with respect to the FAABOS functional categories. 
\end{enumerate}

The rest of the paper is organised as follows: Section II details about the proposed EMAHA-DB1 dataset; Section III presents experiments in machine learning frameworks; Section IV demonstrates the experimental results; and Section V provides a conclusion along with the future scope.

\section{EMAHA-DB1: Proposed sEMG Dataset}	 

\begin{table}[!t]
\caption{List of hand activities}
\label{tab:hand activities with label info}
\setlength{\tabcolsep}{3pt}

\begin{tabular}{p{0.2\columnwidth} |p{0.75\columnwidth} }
\hline
\textbf{Activity No.}                        & \textbf{Activity description }                                                                                                                                       \\
\hline
A0         &  Hand at rest (sitting)
\\
A1         &  Tossing a coin (sitting)
\\

A2                      &      Finger snapping  (sitting)                                                                                       \\
A3        &                    Pulling an empty draw - Posterior view (sitting)                                                                         \\
A4                    & Pulling a draw with weight ($2$kg) - Posterior view      (sitting)                                                                                                                     \\
A5                  &     Pulling an empty draw - Anterior view   (sitting) \\                                                                      A6                      &      Pulling a draw with weight ($2$kg) - Anterior view    (sitting)                                                                                     \\
A7        &                    Pushing an empty draw - Posterior view (sitting)                                                                         \\
A8                    & Pushing a draw with weight ($2$kg) - Posterior view   (sitting)                                                                                                                         \\
A9                  &     Clasping both hands (sitting) \\   
A10                      &      Hand clapping  (sitting)                                                                                       \\
A11        &                    Grasping and holding $1$L water bottle (sitting)                                                                        \\
A12                    & Grasping and holding small hammer (sitting)                                                                                                     \\
A13                  &    Grasping and holding small saw (sitting) \\ 
A14                      &      Writing the phrase "Bio signal lab" on paper with pen - lateral grasp  (sitting)                                                                                       \\
A15        &                 Writing the phrase "Bio signal lab" on board with marker - lateral grasp  (standing)                                                                         \\
A16                    & Lifting a small bucket with $4$L water    (standing)                                                                                                   \\
A17                  &     Typing the phrase "Bio signal lab" on keyboard using single finger (sitting)
 \\
                                                 
A18       &                 Drinking tea/water from a cup - lateral grasp (sitting)                                                                          \\
A19                    & Picking up the phone, placing it to his/her ear and hanging up the phone on table (sitting)                                                                                                      \\
A20                  &     Grasping and holding a book (sitting)
 \\
A21                  &     Grasping and holding a tennis ball (sitting)
 \\
\hline
\end{tabular}
\end{table}

\begin{table}[!t]
\caption{Sensor placement on hand muscle}
\label{tab:Sensor locations}
\setlength{\tabcolsep}{3pt}

\begin{tabular}{p{0.2\columnwidth} |p{0.2\columnwidth} |p{0.5\columnwidth}}
\hline
\textbf{Channel No.}                        & \textbf{Sensor No. }                                                                              & \textbf{Hand muscle name}                                                            \\
\hline

1         &  21621 & Brachio Radialis (BR) muscle
\\
2                      &      21623                                                                                 & Flexor Carpi Radialis(FCR) muscle                                                            \\
3        &                    21624                                                                          & Flexor Carpi Ulnaris (FCU) muscle    \\
4                    & 21625                                                                           & Biceps Brachii (BB) muscle                                                      \\
5                  &     21626                                                                                         & Abductor Pollicis Brevis (APB) muscle     
 \\

\hline
\end{tabular}
\end{table}

\subsection{Data Collection}
\subsubsection{Study participants}
The institutional ethics committee of Indian Institute of Information Technology Sri City (No. IIITS/EC/2022/01) approved the proposed data collection protocol developed in general accordance with the declaration of Helsinki and specific accordance with the “National Ethical Guidelines for Biomedical and Health Research involving human participants" of India.
Twenty-five healthy subjects with no history of upper limb pathology, including $22$ males and $3$ females, participated in the sEMG data collection process. The average age is $28 \pm 6$ years. Before the first session of activities, each of the participants gave written informed consent and the data collection process is completely non-invasive.

\subsubsection{Experimental setup and acquisition protocol} 
The $22$ activities performed by each subject are listed in Table \ref{tab:hand activities with label info}. Each of the hand muscle activity is recorded with a 5-channel Noraxon Ultium wireless sEMG sensor setup \cite{Noraxon} as shown in Fig. \ref{fig:Block diagram}. Five self-adhesive Ag/AgCL dual electrodes were placed at the centre of the five most representative muscle sites of the right arm as shown in Fig. \ref{fig:Block diagram}. Each subject is instructed to sit comfortably with one elbow resting on a table and an arm flexed 90$^{\circ}$ compared to the forearm. The muscle locations are selected according to the atlas in chapter 17 \cite{criswell2010cram} and is given in Table \ref{tab:Sensor locations}.  At the beginning of each session, the participant's hands are cleaned with an alcohol based wet wipe.

Prior to each session, the subject is acquainted with the experiment protocol including a video demonstration of the proposed activities. The total duration of each session is up-to one hour per subject depending on adaptability. Each activity is performed for a maximum duration of $10$s and repeated $10$ times. There is a rest period of $5$s between each of the repetitions and a $30$s gap between the sessions of different activities. Each of the activities consists of two phases: (1) an action and (2) rest. However, some of the activities included an extra release phase. During the action phase, the subject performs the corresponding activity; during the release phase, the subject transitions from the action state to rest  state; and during the rest phase, the subject completely relaxes each of his/her muscles. The time duration for each activity is given in Table \ref{tab:hand activities with timings}, where $T_{X}$, $T_{A}$, $T_{R}$, and $T_{T}$ are the rest, action, release, and total duration, respectively.


\begin{table}[!t]
\caption{Phase-wise durations of each activity.}
\label{tab:hand activities with timings}
\setlength{\tabcolsep}{3pt}
\begin{tabular}{ p{0.05\columnwidth} |p{0.038\columnwidth}| p{0.038\columnwidth} |p{0.038\columnwidth} |p{0.038\columnwidth} || p{0.05\columnwidth} |p{0.038\columnwidth}| p{0.038\columnwidth} |p{0.038\columnwidth} |p{0.038\columnwidth} || p{0.05\columnwidth} |p{0.038\columnwidth}| p{0.038\columnwidth} |p{0.038\columnwidth} |p{0.038\columnwidth}}
\hline
\textbf{No.}                        &  \textbf{$T_{X}$}  & \textbf{$T_{A}$}    & \textbf{$T_{R}$}   & \textbf{$T_{T}$} & \textbf{No.}                        &  \textbf{$T_{X}$}  & \textbf{$T_{A}$}    & \textbf{$T_{R}$}   & \textbf{$T_{T}$} & \textbf{No.}                        &  \textbf{$T_{X}$}  & \textbf{$T_{A}$}    & \textbf{$T_{R}$}   & \textbf{$T_{T}$}           \\ \hline
A1 & 3 & 5 & 0 & 8 & A8 & 3 & 5 & 0 & 8 & A15 & 3 & 15 & 2 & 20\\
A2 & 3 & 5 & 0 & 8 & A9 & 3 & 5 & 0 & 8 & A16 & 5 & 5 & 3 & 13\\
A3 & 3 & 5 & 0 & 8 & A10 & 3 & 5 & 0 & 8 & A17 & 3 & 10 & 2 & 15\\
A4 & 3 & 5 & 0 & 8 & A11 & 3 & 5 & 3 & 11 & A18 & 5 & 5 & 3 & 13\\
A5 & 3 & 5 & 0 & 8 & A12 & 3 & 5 & 3 & 11 & A19 & 5 & 5 & 3 & 13\\
A6 & 3 & 5 & 0 & 8 & A13 & 3 & 5 & 3 & 11 & A20 & 5 & 5 & 3 & 13\\
A7 & 3 & 5 & 0 & 8 & A14 & 3 & 10 & 2 & 15 & A21 & 5 & 5 & 3 & 13\\
\hline
\end{tabular}
\end{table}



\begin{table*}[!t]
\caption{Comparisons of basic data characteristics with benchmark datasets}
\label{tab:stats comparison}
\begin{tabular}{p{0.13\columnwidth} | p{0.3\columnwidth} | p{0.18\columnwidth}| p{0.13\columnwidth} |p{0.14\columnwidth} |p{0.13\columnwidth} | p{0.14\columnwidth} | p{0.1\columnwidth}| p{0.1\columnwidth} | p{0.1\columnwidth} | p{0.1\columnwidth}| p{0.1\columnwidth}}

\hline
\textbf{Dataset Name}                        & \textbf{Action categories} & \textbf{Sensor} & \textbf{No. of Subjects (S)} & \textbf{No. of activities ($N_{A}$) (including rest)}  & \textbf{No. of channels ($N_{c}$)} & \textbf{Sampling frequency ($N_{s}$) (samples per sec)} & \textbf{Rest duration ($T_{X}$)(s)}  & \textbf{Action duration ($T_{A}$)(s)}    & \textbf{Release duration ($T_{R}$)(s)}   & \textbf{No. of repetitions ($N_{R}$)} & \textbf{Total no. of patterns ($N$)}                                                                                                                                                                    \\
\hline

NinaPro DB1 \cite{Atzori2014} &  Gestures, Wrist movements, and Grasping Objects & Otto Bock & 27 & 53 & 10 & 100 & 3 & 5 & -  & 10 & 14310
\\

NinaPro DB2 \cite{Atzori2014} &  Gestures, Wrist movements, Grasping Objects, and Finger pressing movements & Delsys Trigno wireless & 40 & 50 & 12 & 2000 & 3 & 5 & -  & 6 & 12000
\\
NinaPro DB4 \cite{10.1371/journal.pone.0186132} &  Gestures, Wrist movements, and Grasping Objects & Cometa MiniWave & 10 & 53 & 12 & 2000 & 3 & 5 & - & 6 & 3180
\\
BioPatRec DB2 \cite{Ortiz-Catalan2013} &  Gestures, and Wrist and hand movements & Thalmic myoarm band & 17 & 27 & 8 & 2000 & 3 & 3 & - & 3 & 1377
\\
UCI Gesture  \cite{lobov2018latent} &  Wrist and hand movements & Myo Thalmic bracelet & 36 & 7 & 8 & 1000 & 3 & 3 & - & 4 & 1008
\\
Rami-kushaba DB6 \cite{khushaba2014towards} & Hand movements & \raggedright Delsys DE 2.x series EMG sensors & 11 & 40 & 7 & 4000 & 3-5 & 5 & - & 6 & 2640
\\
\textbf{EMAHA-DB1 (Our dataset)} &\raggedright \textbf{Daily activities - Grasping and holding, writing, and draw open/close} & \textbf{Noraxon Ultium sEMG sensor} & \textbf{25} & \textbf{22} & \textbf{5} & \textbf{2000} & \textbf{3-5} & \textbf{5-15} & \textbf{3-5} & \textbf{10} & \textbf{5500}
\\

\hline
\end{tabular}
\end{table*}

\subsubsection{Comparisons with existing datasets }
The characteristics of EMAHA-DB1 data are compared against those of existing sEMG hand activity datasets in TABLE \ref{tab:stats comparison}. Apart from those mentioned in salient features in Introduction, a few additional and distinct characteristics of the EMAHA-DB1 are: 1) the experiments are designed such that hand activities performed consists of three phases of action (contraction/relaxation of muscles), release (retreating of action), and rest (relaxing of muscles), 2) the measurements are acquired with a minimal number of sensors hence requires lower computational resources compared to the existing datasets.

\subsection{Data Preparation}
\subsubsection{Activity segmentation}
\begin{figure}[!t]
\centering
\includegraphics*[width=1.0\columnwidth]
{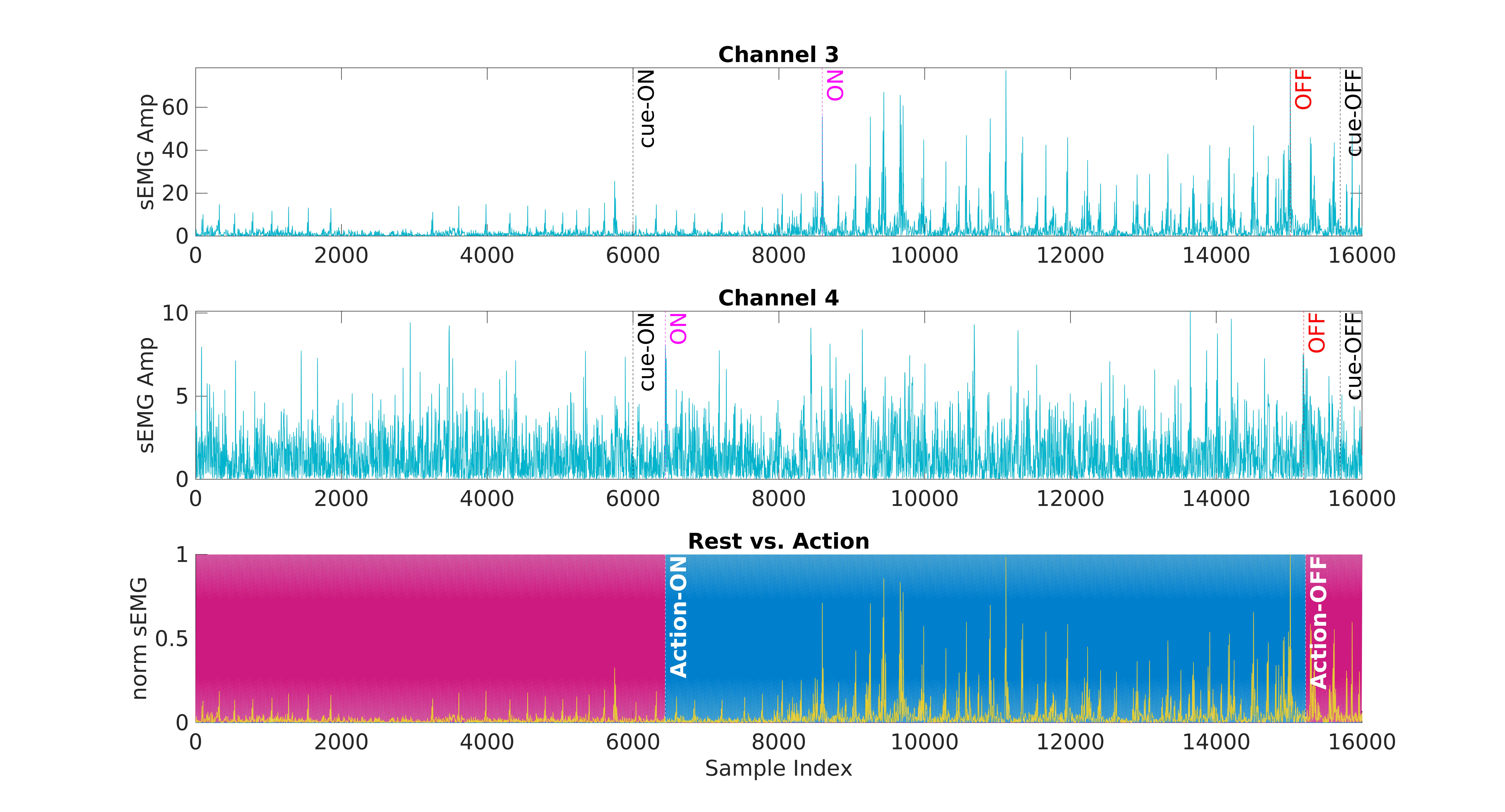}
\caption{Illustration of manual segmentation of sEMG signals for a trial of ADL} 
\label{fig:action_vs_rest_labelling}
\end{figure}

For the sEMG signals in EMAHA-DB1, the preliminary annotations for onset and offset of the actions are performed based on the respective durations of action phases shown in Table \ref{tab:hand activities with timings}. To improve the quality of  activity labels, based on the procedure developed in \cite{malevsevic2021database}, an improved signal segmentation process (listed below) is implemented: 
\begin{enumerate}
    \item Initially, for each trial of each activity performed by each subject, the multi-channel signal is rectified.
    \item For each of these trials, the maximum and minimum values are identified to determine the range $R$. 
    \item The signal strengths at  $R/\sqrt{2}$ ($3$dB amplitude) are considered the thresholds on either side. 
    \item The first signal strength, past the preliminary onset, crossing the $3$dB threshold is identified for each channel. The earliest location among the threshold crossings from the five channels is considered the onset of action.
    \item The trial data is parsed backwards from the end of the action. The first point from the end i.e., the final $3$dB crossing is identified for each channel. The right most location among the crossings from these channels is labelled the offset of action.
    \item Finally, the signal samples between the onset and the offsets are annotated as the action and assigned the corresponding activity number, and the remaining signal is considered to be rest state.
\end{enumerate}
The above procedure is illustrated in Fig. \ref{fig:action_vs_rest_labelling} for a single trial of ADL. It is observed that signal segmentation improves the annotation process of activity vs. rest  which further improves veracity of the classification process.

\subsubsection{FAABOS categories} 
The EMAHA-DB1 is mapped according to function arm activity behavioral observation system (FAABOS) \cite{uswatte2009behavioral,8007300}. Specifically, actions in the EMAHA-DB1 are reorganized into the following five major groups: 1) No object action, 2) object holding, 3) object grasping, 4) Flexion and Extension of Fingers, and 5) writing. The action categories that are mapped into these groups are listed in Table \ref{tab:activity groups}.

\begin{table}[!t]
\caption{FAABOS groups of activities.}
\label{tab:activity groups}
\setlength{\tabcolsep}{3pt}

\begin{tabular}{p{0.18\columnwidth} |p{0.28\columnwidth} |p{0.4\columnwidth}}
\hline
\textbf{Group label}                        & \textbf{Group Name }                                                                              & \textbf{Activity No.}                                                            \\
\hline

0         &  Rest & A0
\\

1                      &   No object action                                                                                  & A2, A9 and A10                                                            \\
2       &                   Hold object                                                                         & A3, A4, A5, A6, A7 and A8    \\
3                    & Object grasping                                                                          & A11, A12, A13, A16, A18, A19, A20 and A21                                                     \\
4                 &     Flexion and Extension of Fingers                                                                                         & A1 and A17     
 \\
 5                 &     Writing                                                                                        & A14 and A15
 \\

\hline
\end{tabular}
\end{table}


\section{Methodology}

\subsection{Problem Statement}
The total number of sEMG patterns in the EMAHA-DB1 is $N = S \times N_{A} \times N_{R}$, where $S$ is the total number of subjects, $N_{A}$ is the number of different activities, and $N_{R}$ corresponds to the number of repetitions per action per subject. The proposed sEMG dataset can be represented as: 
\begin{equation}   \label{eq1}
    \vx = \{\vx_{n}\}_{n=1}^N
\end{equation}
where each observation array $\vx_n$ consists of multiple channels as:
\begin{equation} \label{Sig_Chan_all}
    \vx_n = \{\vx_{n,m}\}_{m=1}^{N_{C}},  ~~~ n = 1, \cdots, N
   \end{equation}
where $N_{C}$ is the number of channels (from different electrodes) and each of these channels consists of an array as:
\begin{equation} \label{sig_Chan_m}
    \vx_{n,m} = \{x_{n,m}(i)\}_{i=1}^{N_T} 
 \end{equation}
where $N_T = N_s \times T_T$ is the number of values in one trial of duration $T_T$ and $N_s$ is the sampling rate (samples/sec). For a given trial, for feature extraction purposes, the signal is divided into $N_{seg}$ segments. Each segment $s_g$ consists of an array as: \begin{equation} \label{sig_Chan_m}
    \vs^j_{g} = \{x_{n,m}(i)\}_{i=1}^{N_g} ~~~ j = 1, \cdots N_{seg} 
 \end{equation}
where $N_g$ is the number of samples in one segment such that $N_{T} = N_{seg} \times N_g$.

\begin{table}[!t]
\caption{Summary of extracted features}
\label{tab:features}
\begin{tabular}{p{0.1\columnwidth} |p{0.6\columnwidth} |p{0.22\columnwidth}}
\hline
\textbf{Feature Set}                        & \textbf{Features }                                                                              & \textbf{Feature Length}                                                            \\
\hline

$F_0$ \cite{karnam2021classification}        &  Mean Absolute Value (MAV), Temporal Spectral Energies (TSE) and Spectral Band Energies (SBE) & $1 \times N_C$, $4 \times N_C$, and $4 \times N_C$   
\\

$F_1$ \cite{young2012classification}                     &      MAV, Zero Crossings (ZC), Slope Changes (SC), and Wavelength (WL)                                                                                 & $1 \times N_C$, $1 \times N_C$, $1 \times N_C$, and $1 \times N_C$                  \\
$F_2$  \cite{geethanjali2014low}     &                   F1 and Auto Regression Coefficients (ARC)                                                                         & $9 \times N_C$ and $2 \times N_C$     \\
$F_3$  \cite{waris2018multiday}                  & F1, Myopulse Rate (MPR), Willison Amplitude (WAMP), and Cardinality                                                                           & $9 \times N_C$, $1 \times N_C$, $1 \times N_C$, and $1 \times N_C$                                                    \\
$F_4$   \cite{al2015improving}               &    Log moments in frequency domain (LMF)                                                                                        & $5 \times N_C$    
 \\
 $F_5$  \cite{turlapaty2019feature}                &     F4, modified LMF, Time domain statistics (TDS), Spectral Band Powers (SBP), Max channel cross correlations, and Local Binary Patterns (LBP)                                                                                        & $5 \times N_C$, $10 \times N_C$, $4 \times N_C$, $4 \times N_C$, $2 \times N_C$, and $2 \times N_C$    
 \\
$F_6$  \cite{campbell2020current}                &     Root Mean Square (RMS), Time Dependent Power spectrum Descriptors (TD-PSD) \cite {al2015improving}, Difference Absolute Standard Deviation Value (DASDV), and Difference Absolute Mean Value (DAMV)                                                                                        & $1 \times N_C$, $6 \times N_C$, $1 \times N_C$, and $1 \times N_C$    
 \\

\hline
\end{tabular}
\end{table}

\begin{figure*}[!t]
\centering
\begin{subfigure}[t]{0.32\textwidth} 
    \includegraphics*[width=1.0\columnwidth]{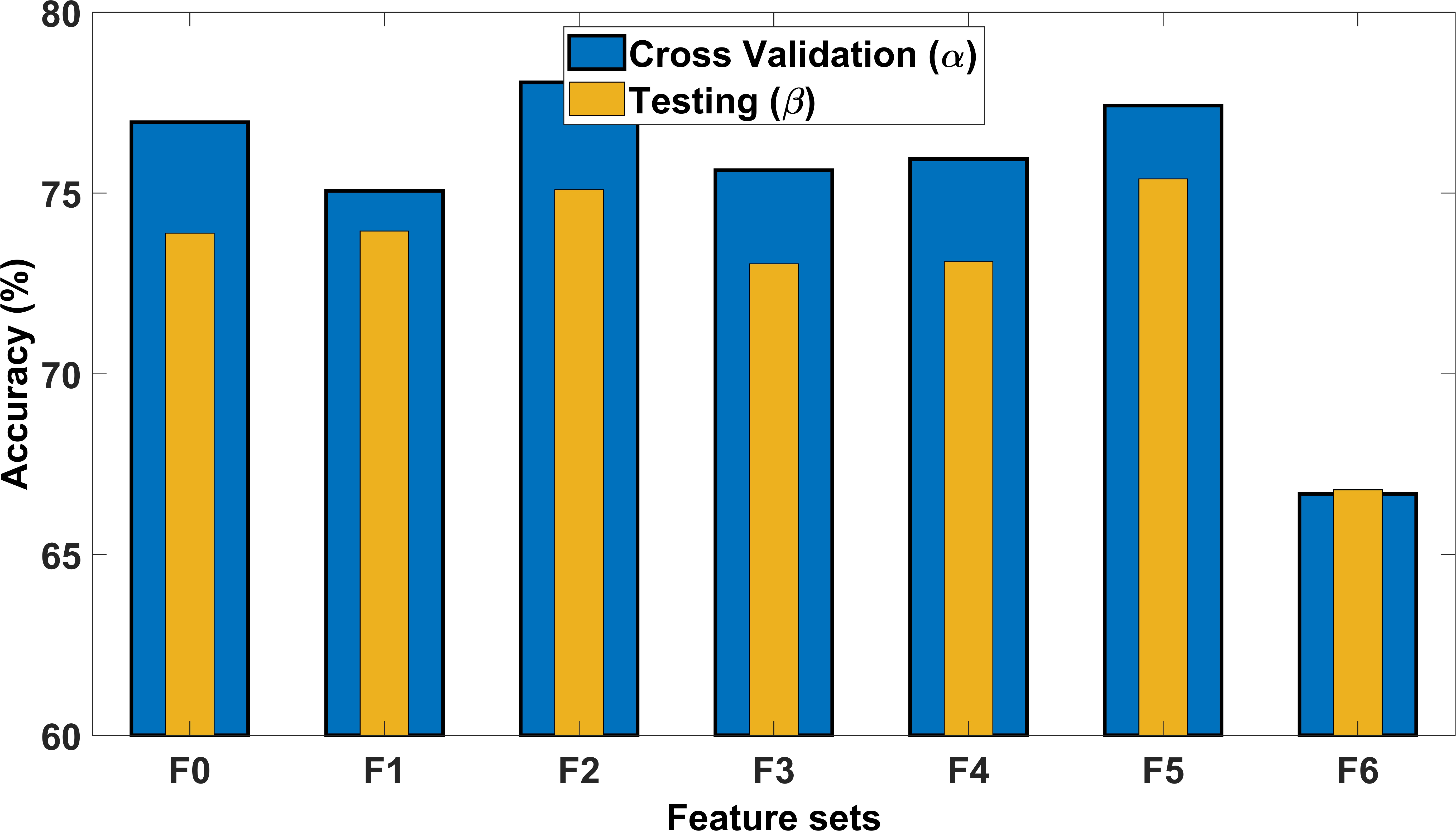} 
    \caption{}
    \label{fig:feat_comp}
\end{subfigure}%
    ~ 
\begin{subfigure}[t]{0.32\textwidth}
    \includegraphics*[width=1.0\columnwidth]{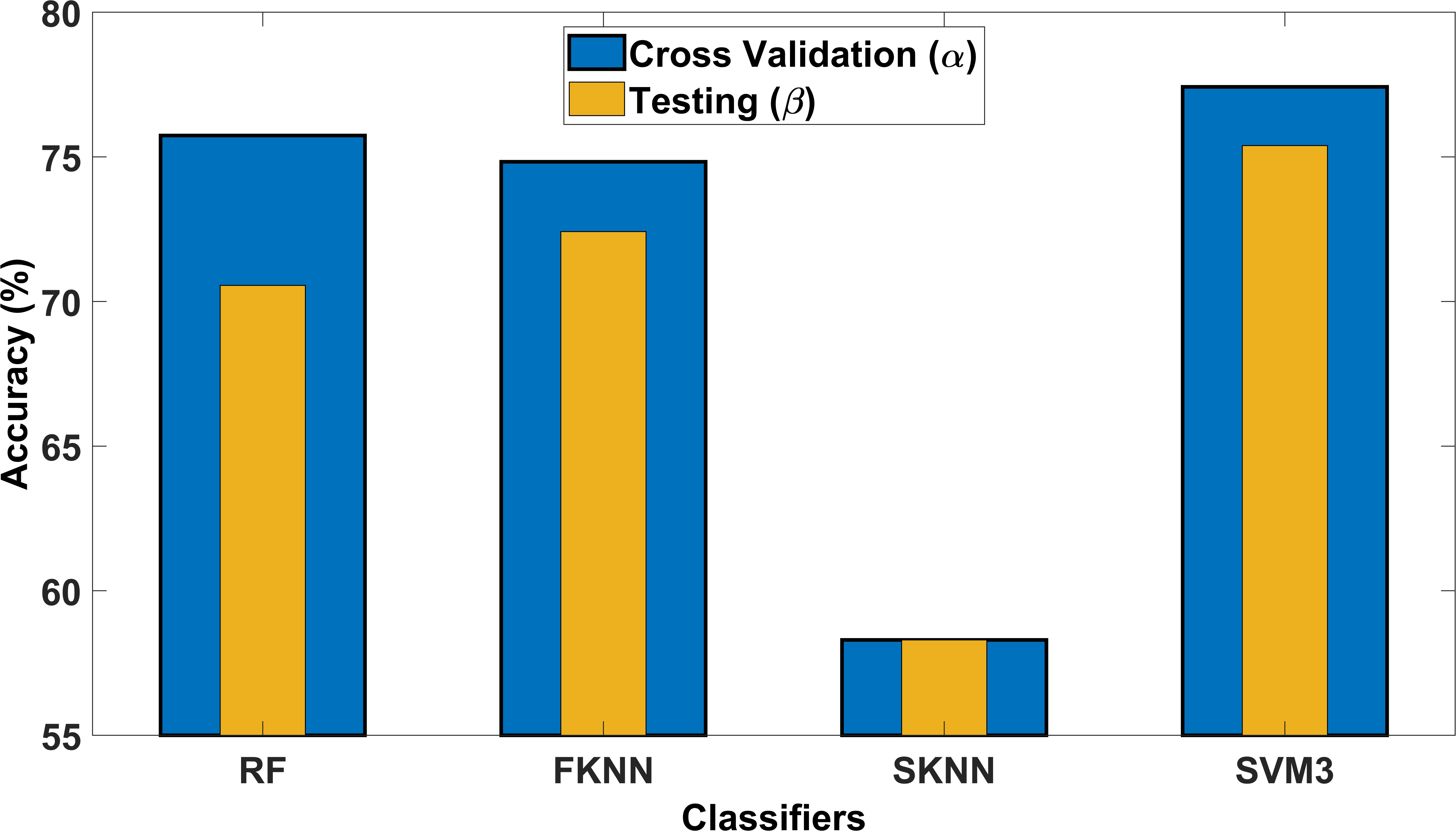}
    \caption{}
    \label{fig:class_comp}
\end{subfigure}%
    ~ 
\begin{subfigure}[t]{0.32\textwidth}
    \includegraphics*[width=1.0\columnwidth]{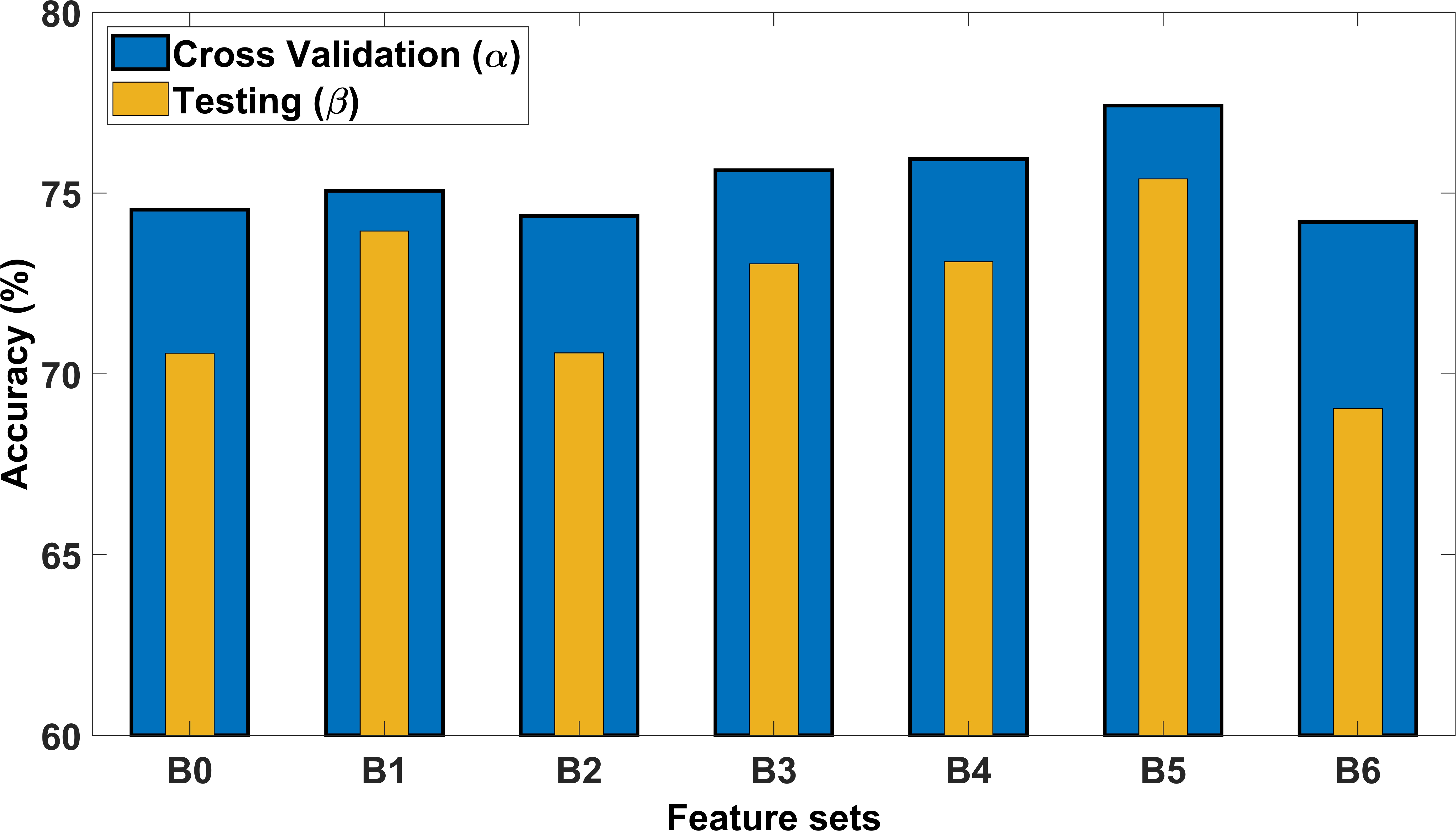}
    \caption{}
    \label{fig:frame_work}
\end{subfigure}
\begin{subfigure}[t]{0.33\textwidth}
    \includegraphics*[width=1.0\columnwidth]{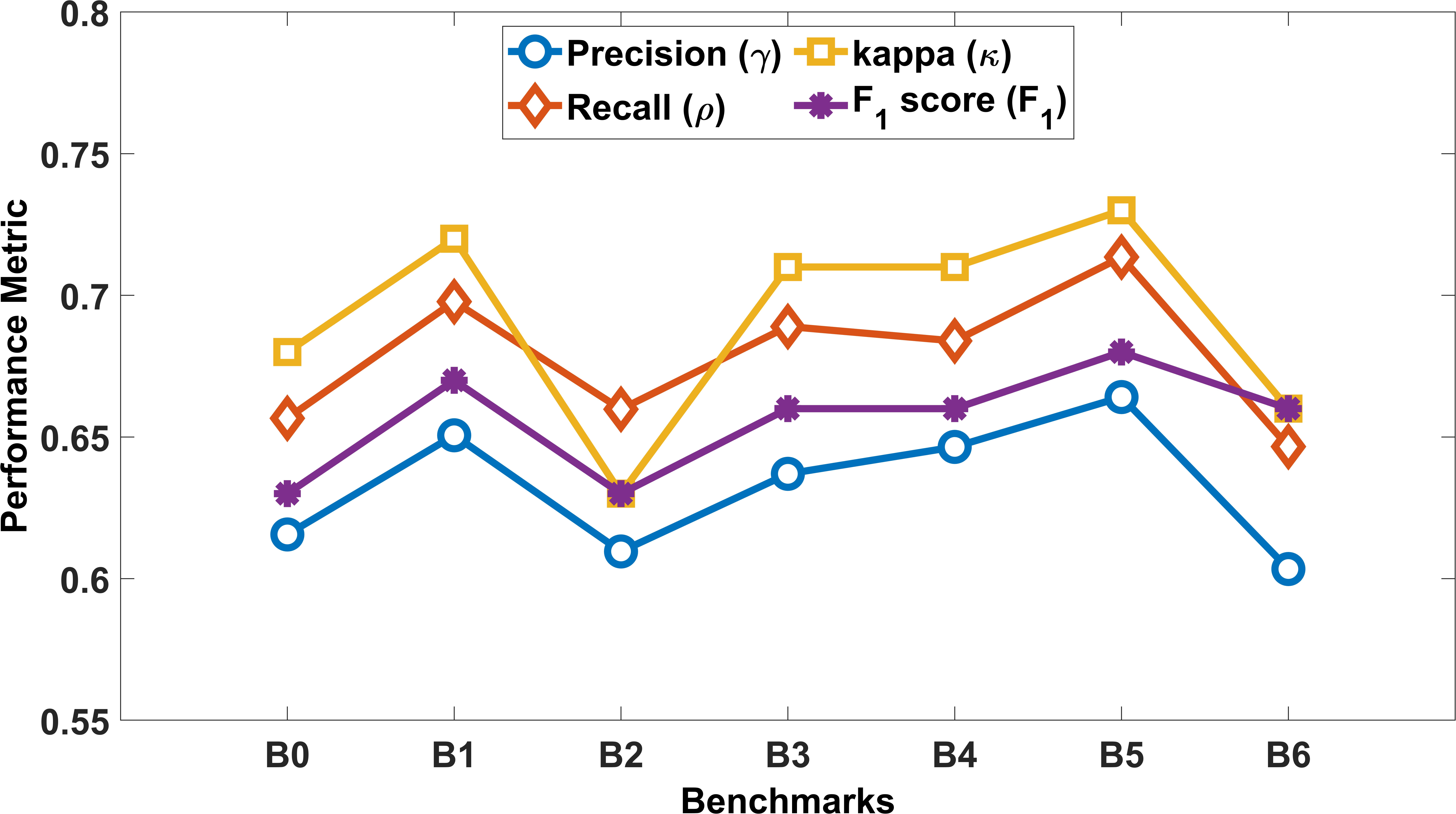}
    \caption{}
    \label{fig:metrics}
\end{subfigure}%
    ~ 
\begin{subfigure}[t]{0.33\textwidth}
    \includegraphics*[width=1.0\columnwidth]{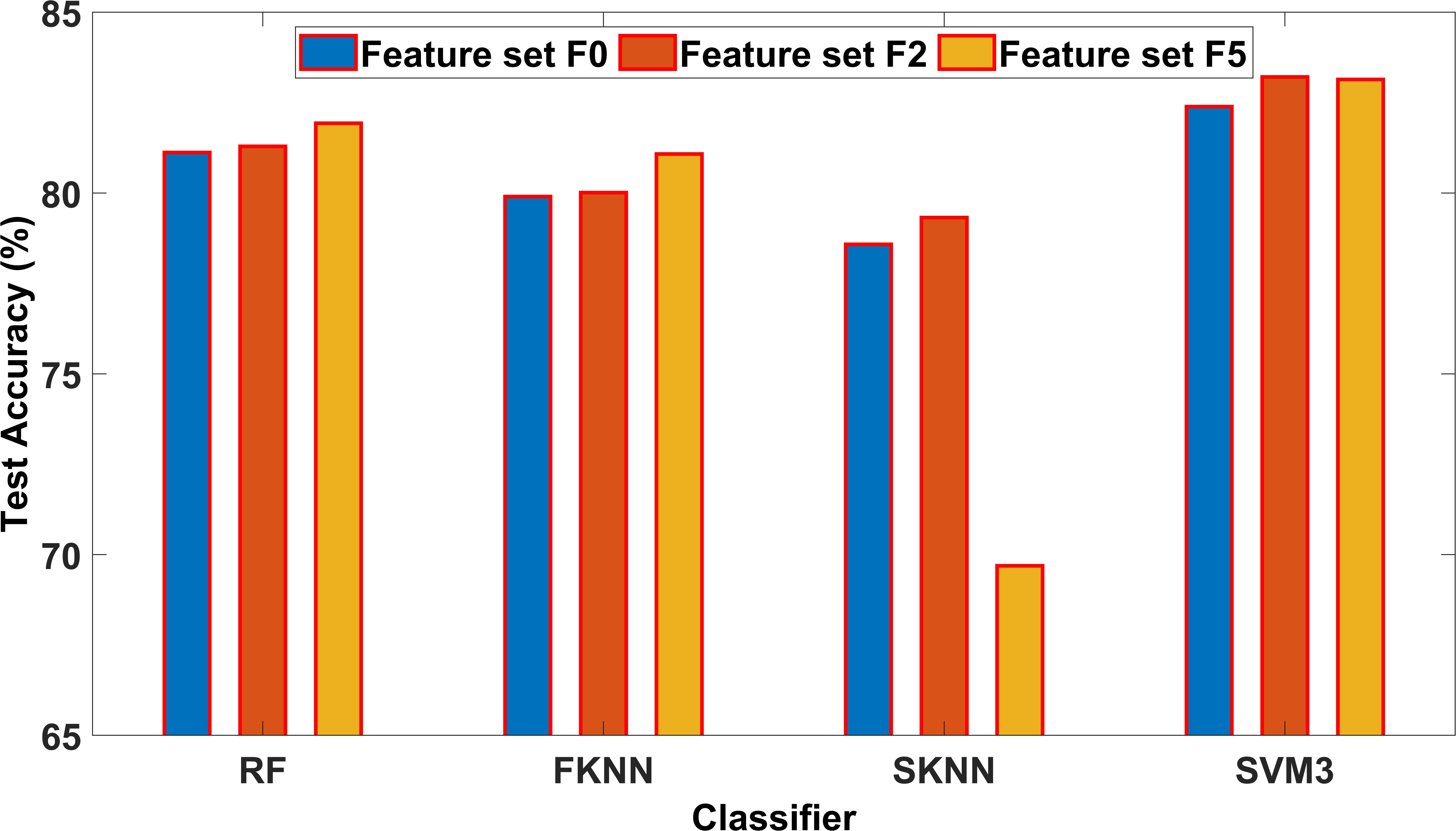}
    \caption{}
    \label{fig:FAABOS categories classifier analysis}
\end{subfigure}
\caption{Performance comparison (a) with different Feature Ensembles with Cubic SVM  (Polynomial SVM of order 3), (b) with different classifiers for the benchmark feature ensemble $F_5$, (c) against benchmark frameworks, (d) against benchmark frameworks in terms of various metrics, and (e) against FAABOS categories frameworks.} 
\label{fig:results}
\end{figure*}

The objective of this study is to map the segmented sEMG signals to the corresponding activity labels (i.e., $\vt_g$ - targets), which can be formulated as:
\begin{equation} 
    f\{\vs_{g}\} \rightarrow \vt_{g}
\label{eq:semg_to_activity}
 \end{equation}
where $\vt_g$ denotes targets  (group labels) as specified in TABLE \ref{tab:activity groups} or individual activity labels as specified in TABLE \ref{tab:hand activities with label info}. The mapping function in (\ref{eq:semg_to_activity}) is implemented by a supervised classifier. For the mapping function, appropriate features are required that represent the underlying inverse kinematic relationships between the sEMG signals and the corresponding activity performed. 

\begin{table}[!t]
\caption{Numerical setup for classifiers.}
\label{tab:setup classifiers}
\begin{tabular}{p{0.2\columnwidth} |p{0.75\columnwidth}}
\hline
\textbf{Classifier}                        & \textbf{Model Setup}                                                                                              \\
\hline

Fine KNN        & No.of neighbours = $5$, Distance Metric = Cityblock,
Distance weight = Squared Inverse
\\

Ensemble KNN                      &   No.of learning cycles = $30$, learners = KNN, Subspace dimension = $25$                                                          \\
Cubic SVM        &                   Polynomial kernel, Order = $3$, Box constraint = $1$, Multi-class Method = one-vs-one    \\
Random Forest                  & No.of bags for bootstrapping = $300$                    \\

\hline
\end{tabular}
\end{table}

\subsection{Feature Extraction}
In this work, the following feature sets are adapted from \cite{karnam2021classification}: $F_0$, $F_1$, $F_2$, $F_3$, $F_4$, and $F_5$ with an additional feature set $F_6$ consisting of root mean square (RMS), time dependent power spectrum descriptors (TD-PSD), difference absolute standard deviation value (DASDV), and difference absolute mean value (DAMV). Note the features are computed for each segment and concatenated to build the full feature vector. The extracted feature sets are summarized in Table \ref{tab:features}.

\subsection{Supervised Classifiers}
In this paper, four algorithms including random forest (RF), fine K-nearest neighbour (FKNN), ensemble KNN (sKNN) and cubic support vector machine (SVM3) are considered for sEMG signal classification task. The classifiers are trained and tested with subject-wise data and the average performance is reported. The hyperparameter settings for different machine learning algorithms used in this work are summarized in Table \ref{tab:setup classifiers}. The performance of classifiers is evaluated using the standard metrics such as cross validation accuracy ($\alpha$), testing accuracy ($\beta$), Kappa coefficient ($\kappa$), precision ($\gamma$), recall ($\rho$) and F-1 score ($F_1$).
 

\begin{table}[!t]
\caption{Feature ensemble vs benchmark classifier setup.}
\label{tab:benchmark frameworks}
\centering
\begin{tabular}{p{0.04\columnwidth} |p{0.112\columnwidth} |p{0.11\columnwidth}|p{0.135\columnwidth}||p{0.04\columnwidth} |p{0.112\columnwidth} |p{0.11\columnwidth}|p{0.14\columnwidth}}
\hline
\textbf{FE} & \textbf{FL} & \textbf{BF} & \textbf{Classifier} & \textbf{FE} & \textbf{FL} & \textbf{BF} & \textbf{Classifier}\\
\hline
$F_0$ & $9 \times N_C$ & $B_0$ \cite{karnam2021classification} & Fine KNN & $F_4$ & $5 \times N_C$ & $B_4$ \cite{al2013classification} & SVM3\\
$F_1$ & $4 \times N_C$ & $B_1$ \cite{young2012classification} & SVM3 & $F_5$ & $27 \times N_C$ & $B_5$ \cite{turlapaty2019feature} & SVM3\\
$F_2$ & $11 \times N_C$ & $B_2$ \cite{geethanjali2014low} & Fine KNN & $F_6$ & $9 \times N_C$ & $B_6$ \cite{Atzori2014} & RF\\
$F_3$ & $12 \times N_C$ & $B_3$ \cite{waris2018multiday} & SVM3 & & & &\\
\hline
\end{tabular}
\end{table}









\section{Classification Experiments, Results \& Analysis}
\label{sec:Technical Validation}
The developed EMAHA-DB1 sEMG dataset is analyzed using the state-of-the-art classification and feature extraction methods as detailed below.

\subsection{Pre-processing and Data Split-up}
Based on the procedure described in \cite{karnam2022emghandnet}, the recorded sEMG data is pre-processed as follows. First, the sEMG data is filtered to remove power line noise at $50$Hz. Then a first order Butterworth low pass filter is applied at a cut-off frequency of $500$Hz. Finally, wavelet denoising of order $8$ with the symlet as the mother wavelet is applied. The data from each subject is split trials-wise into $70\%$ for training and $30\%$ for testing as per the splitting method in \cite{karnam2022emghandnet}. A non overlapping moving window segment of $N_g = 200$ samples is considered with duration $T_{seg} = 100ms$. The number of features obtained per segment $s_g$ are summarized in Table \ref{tab:benchmark frameworks}.

\subsection{Experiments}
In this paper, as mentioned earlier two case studies are carried out as follows, 1) classification of individual action categories listed in Table \ref{tab:hand activities with label info}, in this case study, the performance is analyzed with respect to feature ensembles, classifiers, benchmark classification frameworks and finally feature visualization; 2) classification of FAABOS categories listed in Table \ref{tab:activity groups}, in the second case study, the performance is analyzed with respect to feature ensembles followed by an analysis of the most relevant features with respect to the muscle sites. 

\subsection{Case Study 1: Results and Analysis}

\subsubsection{Comparison with feature ensembles}
The feature sets $F_0$-$F_6$ are analyzed in this comparison study. Each of the feature set is utilised as input for SVM3 and their performance metrics $\alpha$ and $\beta$ are evaluated. As shown in Fig. \ref{fig:feat_comp}, the best performance is produced by the feature set $F_5$ ($\alpha = 77.42$ and $\beta = 75.39$). The next best feature ensemble $F_2$ lags behind by $0.3\%$ at $\alpha = 78.06$ and $\beta = 75.09$. The feature ensemble $F_6$ has produced the least classification performance ($\alpha = 66.68$ and $\beta = 66.79$).

\begin{table}[!t]
\caption{Muscle vs action mapping.}
\label{tab:Muscle vs Act Map}
\begin{tabular}{p{0.51\columnwidth} |p{0.44\columnwidth}}
\hline
\textbf{Muscle}                        & \textbf{Major functionality of the muscle}\\
\hline
Biceps Brachii (BB) muscle      & Flexes elbow joint, Supinates forearm and hand at radioulnar joint \\

Brachio Radialis (BR) muscle &  Flexes elbow joint \\
Flexor Carpi Radialis (FCR) muscle        &                   Flexes and abducts hand at wrist    \\
Flexor Carpi Ulnaris (FCU) muscle                  & Flexes and adducts wrist \\
Abductor Pollocis Brevis (APB) muscle                  & Abducts joints of thumb                  \\

\hline
\end{tabular}
\end{table}

\subsubsection{Comparison with classifiers}
In this experiment, the classification performance of the standard machine learning algorithms such as the RF, FKNN, sKNN and SVM3 using the $F_5$ feature set is analyzed. As shown in Fig. \ref{fig:class_comp}, the best performance is produced by the SVM3 classifier ($\alpha = 77.42$ and $\beta = 75.39$) and then by FKNN ($\alpha = 74.83$ and $\beta = 72.42$). The least performance is obtained with SKNN classifier ($\alpha = 58.4$ and $\beta = 58.3$). Thus, it is observed from this experiment that for the feature set $F_5$ the SVM3 classifier outperforms other benchmark classifiers.

\subsubsection{Comparison with benchmark algorithms}
The most suitable classification framework for the  EMAHA-DB1 is determined by comparisons with the existing sEMG benchmark classification methods consisting of respective combinations of a feature ensemble and a classification framework as listed in Table \ref{tab:benchmark frameworks}. The benchmark $B_i$ indicates the classification framework with feature set $F_i$ for $i = 0,1,\cdots,6$. The parameter setups of the different classifiers used in the numerical analysis are also shown in Table \ref{tab:setup classifiers}. The performance of these classifiers is analyzed based on the cross validation accuracy ($\alpha$) and the test accuracy ($\beta$) with the corresponding results shown in Fig. \ref{fig:frame_work}. The benchmark $B_5$ has achieved state-of-the-art performance with $\alpha$ = $77.42$ and $\beta$ = $75.39$. The lowest performance among the compared benchmarks is for $B_6$ with $\alpha$ = $74.2$ and $\beta$ = $69.04$.
The other performance metrics (i.e., $\kappa$, $\gamma$, $\rho$, and $F_1$) of the benchmark frameworks are shown in Fig. \ref{fig:metrics}. The benchmark framework $F_5$ has achieved highest values for each of the performance metrics, i.e., $\kappa$ = $0.73$, $\gamma$=$0.66$, $\rho$=$0.71$, and $F_1$ = $0.68$. The runner-up is $B_6$ framework with metric values $\kappa$ = $0.66$, $\gamma$= $0.60$, $\rho$= $0.64$, and $F_1$ = $0.66$.

\subsubsection{Feature Visualization by t-SNE}
\begin{figure}[!t]
\centering
\begin{subfigure}[t]{0.235\textwidth}
    \includegraphics*[width=1\columnwidth]{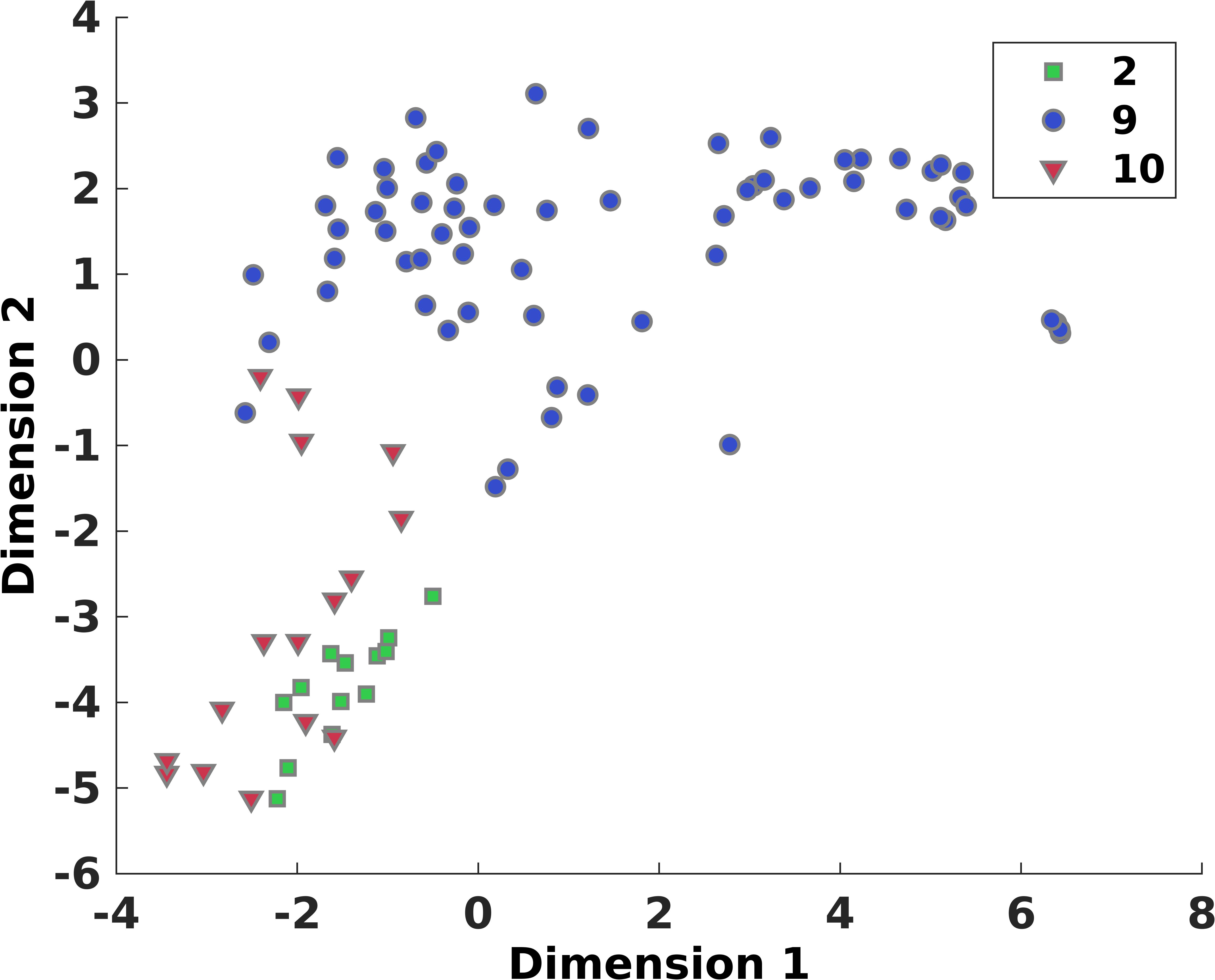} 
    \caption{}
    \label{fig:tsne_group1}
\end{subfigure}%
    ~ 
\begin{subfigure}[t]{0.235\textwidth}
    \includegraphics*[width=1\columnwidth]{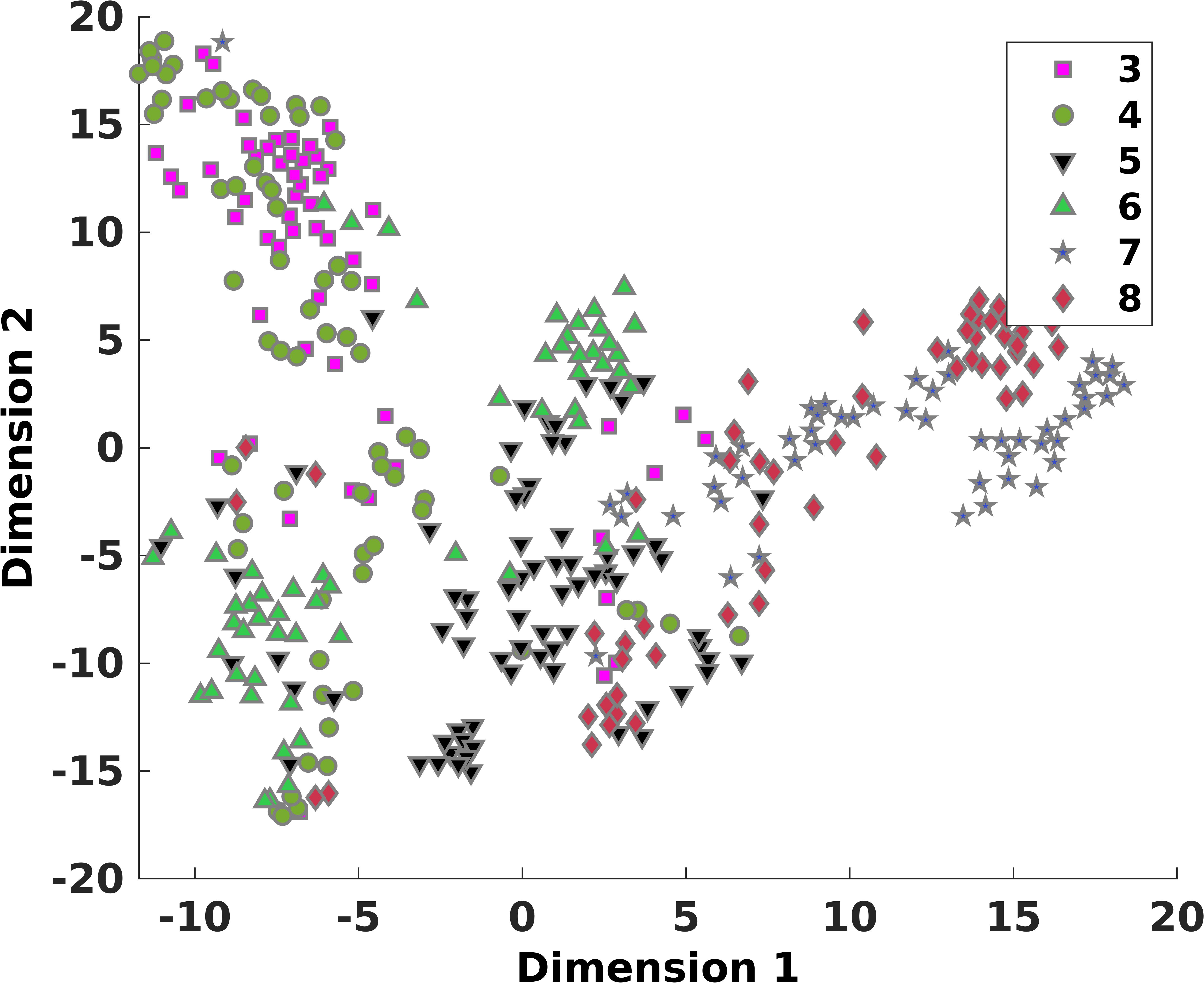} 
    \caption{}
    \label{fig:tsne_group2}    
\end{subfigure}
\begin{subfigure}[t]{0.235\textwidth}
    \includegraphics*[width=1\columnwidth]{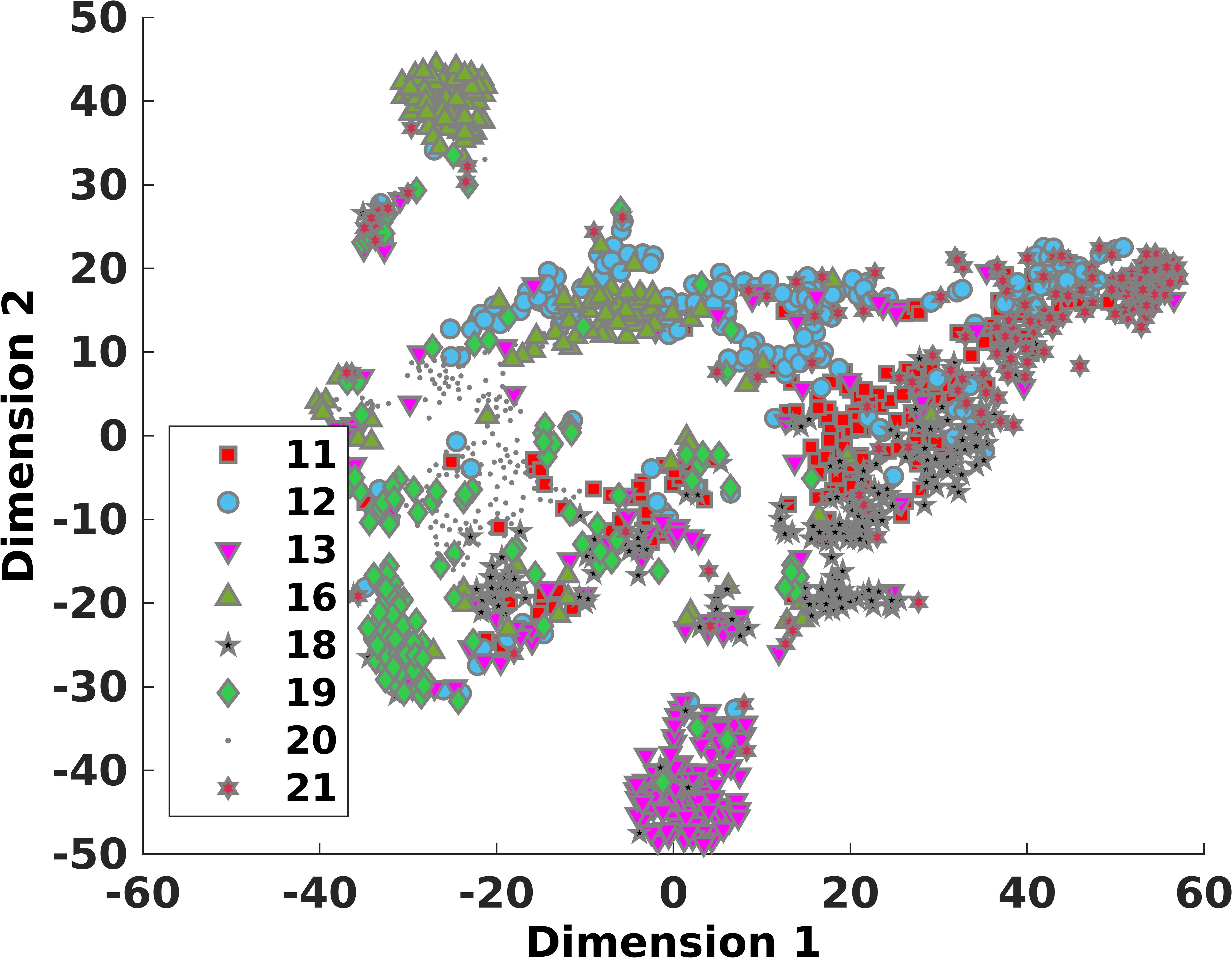} 
    \caption{}
    \label{fig:tsne_group3}    
\end{subfigure}%
    ~ 
\begin{subfigure}[t]{0.235\textwidth}
    \includegraphics*[width=1\columnwidth]{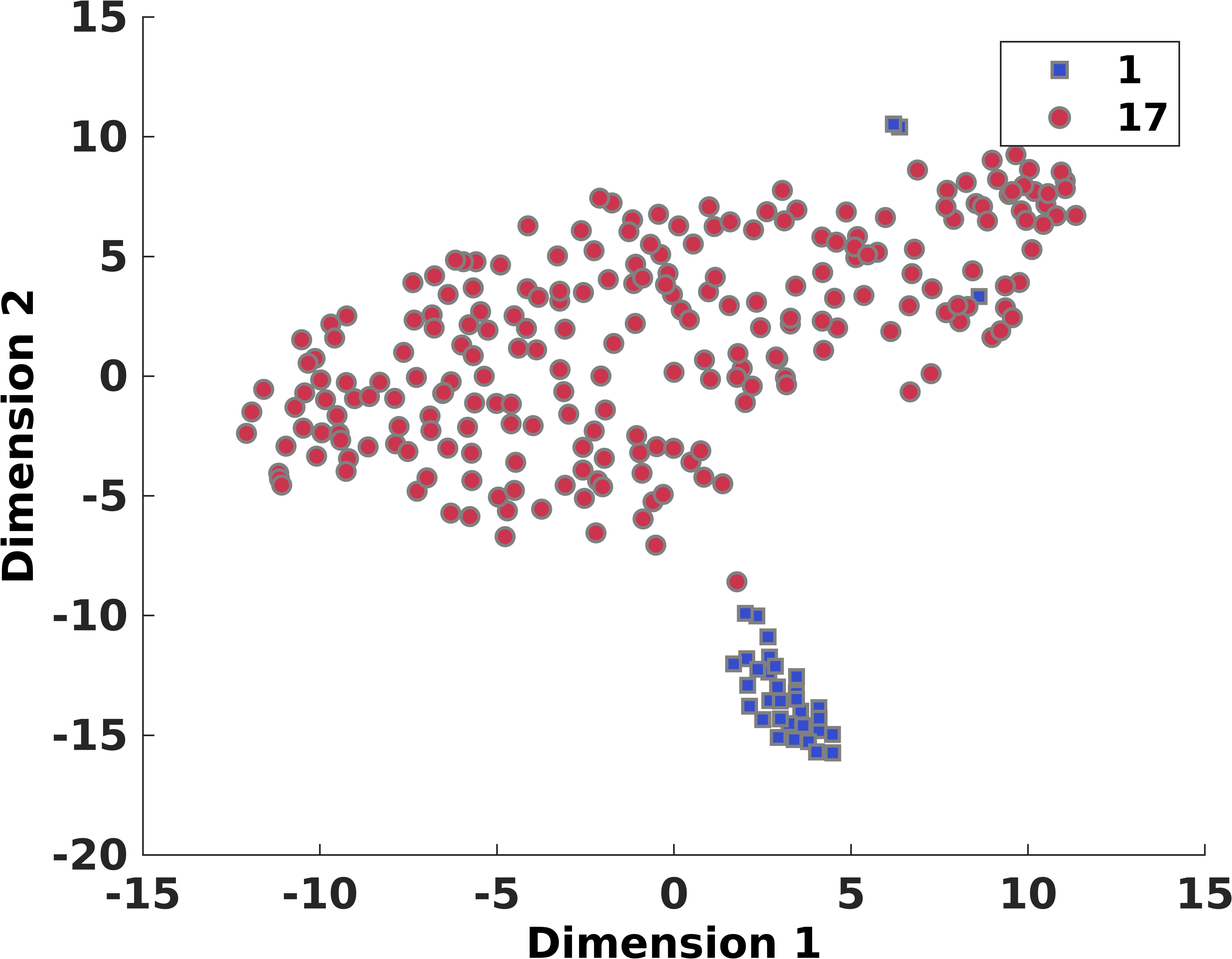} 
    \caption{}
    \label{fig:tsne_group4}    
\end{subfigure}
\begin{subfigure}[t]{0.235\textwidth}
    \includegraphics*[width=1\columnwidth]{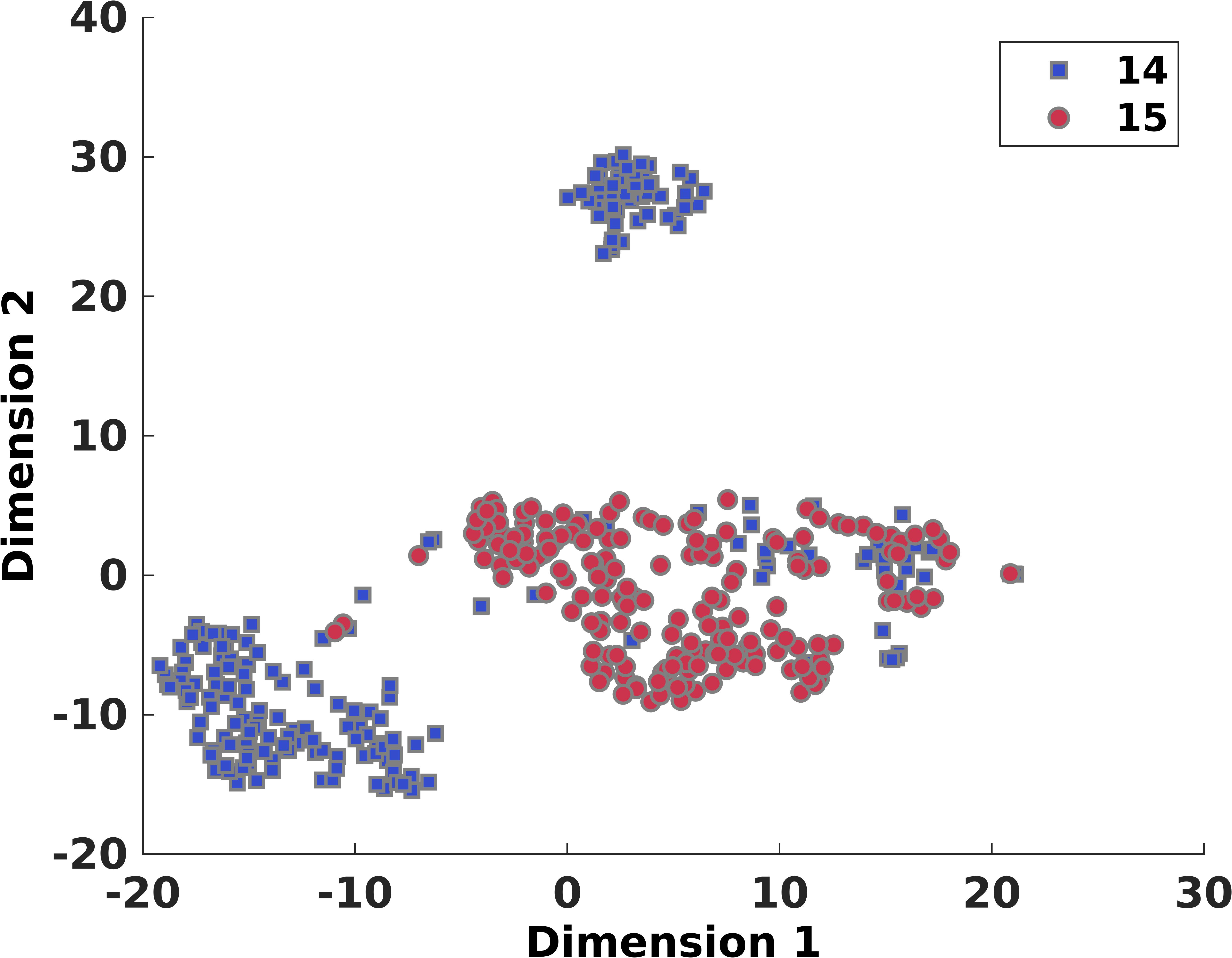} 
    \caption{}
    \label{fig:tsne_group5}    
\end{subfigure}  
\caption{t-SNE plots of feature set for (a) group 1, (b) group 2, (c) group 3, (d) group 4, and (e) group 5, respectively.}
\label{fig:tsne}
\end{figure}

The following analysis is meant for the $22$ individual action categories however carried out FAABOS group wise. 
Among the feature sets $F_0$ to $F_6$, it is observed that $F_5$ is the best performing feature set, hence used for sequential feature selection analysis (SFS). From SFS, the most relevant features for each group of hand activities are identified and further used for analysis with t-distributed Stochastic Neighbourhood Embedding (t-SNE) \cite{van2008visualizing}. 
The top $6$ feature columns of $84$, $85$, $96, 97$, $101$, and $105$ are used in this study. The columns with higher ranking are $84$ and $85$ that correspond to the features of mean and variance respectively (from TDS feature set), and $96, 97, 101,$ and $105$ that correspond to the spectral bands [$0$ $(N_s/8)$] and [$(N_s/8)$ $(N_s/4)$] of SBP feature set \cite{karnam2021classification}.
The flexion and extension of elbow and wrist flexion and extension are mainly supported by the  muscle groups BB, BR, FCR and FCU \cite{VanDeGraaffKentM.KentMarshall1998Ha/K} as given in TABLE \ref{tab:Muscle vs Act Map}. The action categories in group 2 and group 3 involve the common muscle movements including elbow flexion and extension, wrist flexion and extension and pronation and supination as shown in TABLE \ref{tab:Group vs Act vs Mus Map}. Hence From Fig. \ref{fig:tsne_group2} and \ref{fig:tsne_group3}, the clusters for some of the actions overlap due to involvement of similar muscle groups across actions with same basic muscle movements. The actions within group 1, group 4 and group 5 are clearly separable which can be observed from Fig. \ref{fig:tsne_group1}, Fig. \ref{fig:tsne_group4} and Fig. \ref{fig:tsne_group5}, respectively.

\subsection{Case Study 2: Results and Analysis}

\subsubsection{Comparison of FAABOS categories with feature ensembles}
The sEMG signals from the EMAHA-DB1 are classified based on FAABOS categories specified in Table \ref{tab:activity groups}. The six FAABOS categories of sEMG signals are trained and tested with the top three feature sets such as $F_0$, $F_2$ and $F_5$ and the corresponding results are plotted in Fig. \ref{fig:FAABOS categories classifier analysis}. The best performance is produced by the SVM3 classifier ($\alpha$ = $86.54$ and $\beta$ = $83.21$) with the feature set $F_2$. The next best performance is produced by the same SVM3 classifier ($\alpha$ = $85.85$ and $\beta$ = $83.14$), but with the feature set $F_5$ having a slight variation of $0.07\%$. The least performance is observed for feature set $F_0$ with SKNN classifier ($\alpha$ = $85.66$ and $\beta$ = $82.39$).


\subsubsection{Feature Visualization by t-SNE for FAABOS groups}
\begin{figure}[!t]
\centering
\includegraphics*[width=0.9\columnwidth]
{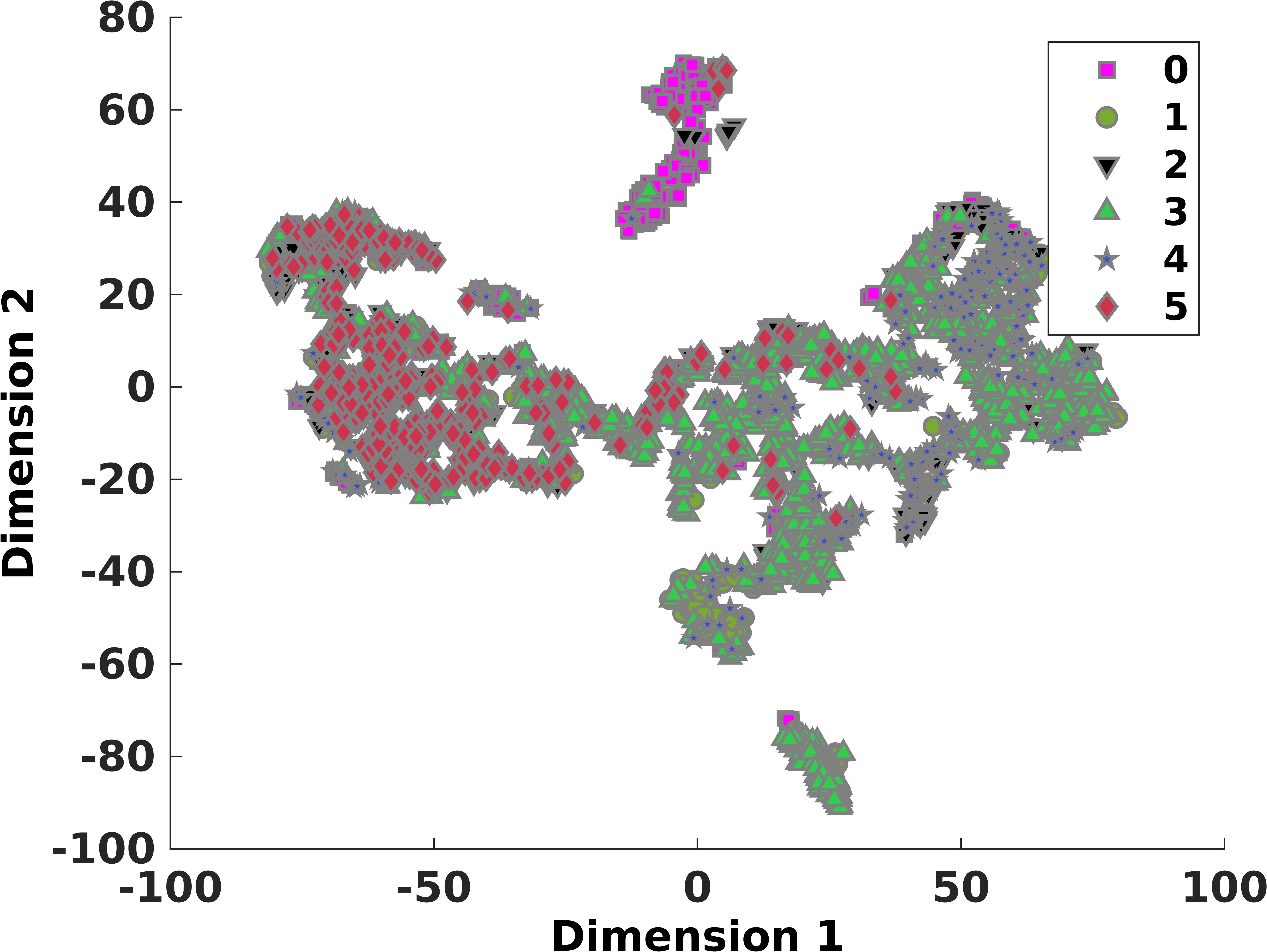}
\caption{t-SNE plot of feature set for six FAABOS groups} 
\label{fig:FAABOS tsne plot}
\end{figure}

This analysis is carried out for $6$ FAABOS categories. Among the feature sets $F_0$, $F_2$, and $F_5$, it is observed that $F_2$ is the best performing feature set and used for SFS analysis. From SFS, the top 6 feature columns $1$, $3$, $4$, $5$, $19$, and $23$ are used in this study. The columns with higher ranking are $1$, $3$, $4$, and $5$ that correspond to mean absolute value (MAV), $19$ corresponds to the waveform length, and $23$ corresponds to auto regressive coefficients. The t-SNE plot is generated with high ranking column features as shown in Fig. \ref{fig:FAABOS tsne plot}. It is observed that the action and rest clusters are clearly separable, but clusters within action groups are overlapping due to similar muscle group involvement. Based on a recent review of sEMG studies of muscle groups and their functions \cite{Jarque-Bou2021}, the muscles FCR, FCU, BR and BB are mapped to the major functions involved in each of the FAABOS categories in our study and detailed in Table \ref{tab:Group vs Act vs Mus Map}.   

\begin{table}[!t]
\caption{FAABOS group vs actions vs muscle mapping.}
\label{tab:Group vs Act vs Mus Map}
\begin{tabular}{p{0.18\columnwidth} |p{0.6\columnwidth}|p{0.15\columnwidth}}
\hline
\textbf{Group}                        & \textbf{Major actions involved}  & \textbf{Muscles}                                                                                                \\
\hline

No object action (1)    & Wrist flexion \& extension and hand digit manipulation & FCR, FCU, BR
\\

Hold object (2)                     &  Elbow flexion \& extension, Wrist flexion \& extension, and Forearm Pronation \& Supination &   BB, BR, FCR, FCU                                                      \\
Object grasping (3)       &                 Elbow flexion \& extension, Wrist flexion \& extension, Forearm Pronation \& Supination, and hand digit manipulation &   FCR, FCU, BR, BB    \\
Flexion and Extension of Fingers (4)                 & Wrist flexion \& extension and hand digit manipulation &   BB, BR, FCR, FCU   \\
Writing  (5)               & Elbow flexion \& extension, Wrist flexion \& extension, and hand digit manipulation & FCR, BB, FCU, APB             \\

\hline
\end{tabular}
\end{table}

\subsection{Discussion}
The SVM3 method has the best classification performance in case of the FAABOS categories (no. classes  = $5$). This can be explained by relatively less number of classes and ability of feature ensemble $F_5$ to better capture the representation at functional category level. The ML framework's performance may need further improvement. This performance can be explained by relatively higher number of activities and higher intra-class correlations. The feature visualizations with t-SNE has shown better separability of activities within FAABOS groups. A clear separation between rest and action is also observed in t-SNE plot across FAABOS groups.

\section{Conclusion \& Future scope}
\label{sec:Conclusion}
In this paper, we have collected a novel sEMG dataset (EMAHA-DB1) of $22$ activities of daily living from Indian population. The EMAHA-DB1 includes a few activities that are not considered in existing datasets. The sEMG EMAHA-DB1 dataset is compared against the publicly available existing sEMG datasets. The dataset is analyzed from different perspectives including feature set analysis in time domain and frequency domain, individual action classification, FAABOS category classification and feature visualization using t-SNE.
In the above mentioned analysis, the modified LMF, time domain statistical (TDS) feature, spectral band powers (SBP), channel cross correlation and local binary patterns (LBP) ensemble feature set ($F_5$) with Cubic SVM classifier has obtained highest test accuracy of $\beta$ = $75.39\%$. Additionally, in the FAABOS groups classification, the best performance is again produced by the cubic SVM classifier ($\beta = 83.21$) with the feature set consisting of energy features and auto regressive coefficients ($F_2$). Finally, the visual analysis using t-SNE plots showed that the extracted feature set is able to clearly distinguish the ADL activities within a group. The obtained results indicate that the EMAHA-DB1 can be successfully used as a benchmark for the development of hand gesture recognition system, physiological analysis and clinical studies of sEMG for ADL.

In terms of future work, the framework may need further innovation in terms of features to improve the classification performance; the EMAHA-DB1 is analysed using only machine learning classifiers, there is a scope for improvement with deep learning; the dataset can also be analysed by decomposing the time series with wavelets or empirical mode decomposition (EMD) techniques; finally, the EMAHA-DB1 dataset can also be analysed for learning the statistical distributions.

\section*{Acknowledgment}
This research is funded by SERB, Govt. of India under Project Grant No. CRG/2019/003801.

\bibliographystyle{IEEEtran} 
\bibliography{biblio}
 
\end{document}